\documentclass[a4paper,10pt,twoside]{cpc-hepnp}

\usepackage{multicol}
\usepackage{graphicx}
\usepackage{booktabs}
\usepackage{amssymb,bm,mathrsfs,bbm,amscd}
\usepackage[tbtags]{amsmath}
\usepackage{lastpage}

\begin{document}
%\begin{CJK*}{GBK}{song}

%\fancyhead[R]{\footnotesize Submitted to 'Chinese Physics C'}
%\fancyhead[c]{\small Chinese Physics C~~~Vol. 37, No. 1 (2013) 010201} \fancyfoot[C]{\small 010201-\thepage}

\footnotetext[0]{Received 28 Feb 2014}

\title{Optimization of single-step tapering amplitude and energy detuning for high-gain FELs\thanks{partly supported by the Major State Basic Research Development Program (2011CB808301) and National Natural Science Foundation of China (11205156) }}

\author{%
      Li Heting
\quad Jia Qika
} \maketitle

\address{%
National Synchrotron Radiation Laboratory, University of Science and
Technology of China, Hefei, 230029, Anhui, China\\
%$^2$ {\bf Example}: Institute of High Energy Physics, Chinese Academy of Sciences, Beijing 100049, China\\
}

\begin{abstract}
We put forward a method to optimize the step-tapering amplitude of undulator strength and initial energy detuning of electron beam to maximize the saturation power of high gain FELs£¬based on the physics of longitudinal electron beam phase space. Using the FEL simulation code GENESIS, we numerically demonstrate the accuracy of the estimations for parameters corresponding to the linac coherent light source and the Tesla test facility.
\end{abstract}



%\footnotetext[0]{\hspace*{-3mm}\raisebox{0.3ex}{$\scriptstyle\copyright$}2013
%Chinese Physical Society and the Institute of High Energy Physics
%of the Chinese Academy of Sciences and the Institute
%of Modern Physics of the Chinese Academy of Sciences and IOP Publishing Ltd}%

\begin{multicols}{2}

\section{Introduction}

High-gain free-electron lasers (FELs), such as self-amplified spontaneous emission (SASE) and seeded harmonic generation (e.g. HGHG, EEHG), is capable of generating extremely high-brightness radiation in the ultraviolet and X-ray wavelengths. However, for a uniform-parameter undulator, the FEL efficiency at saturation is roughly given by the FEL scaling parameter   [1], where   is typically on the order of 10-3. Therefore, variable parameter undulators are broadly used in FEL operation to enhance the performance [2, 3].

We have investigated the existed tapering strategies. One uses the "standard" KMR formulation [4] in which a synchronous electron with ponderomotive phase equal to the synchronous phase   maintain its resonant energy  throughout the tapered undulator. As shown by KMR, maximizing the product of the area of ponderomotive well and   occurs at  . Thus, the bucket decelerates together with the trapped electrons, yielding more energy in the form of radiation. A self-design taper algorithm based upon the KMR formalism has been implemented in the GINGER simulation code [5]. Another approach is present in Ref. [6], which empirically optimizes K(z) that maximizes the output power at a fixed total undulator length without necessarily trying to keep the trapped particle fraction large at undulator exit.

These tapering strategies are very useful for long undulators and can increase the radiation power by several times even by a few orders of magnitude. In practice, tapering of long undulator is implemented through multiple step-tapering. However, FEL facilities usually construct undulators with the length equal to nominal saturation length or a little longer. On this condition, the FEL power enhancement of single-step tapering can be considerable and not much worse than other tapering schemes.

In this paper, we put forward a new method to optimize the single-step tapering amplitude of undulator strength for high gain FELs, based on the physics of longitudinal electron beam phase space. Then we numerically investigate the energy detuning of electron beam in the same way and develop an empirical formula to optimize the energy detuning based on simulations.

\section{Optimization of single-step tapering }

In a free electron laser, the electron beam and radiation wave interact continually under the resonant condition. For a uniform undulator with undulator period~$\lambda _u$~and undulator strength parameter~$K$~, the fundamental resonant wavelength is
\begin{eqnarray}
\lambda _s  = \frac{{\lambda _u }}{{2\gamma ^2 }}(1 + \frac{{K^2 }}{2})
\end{eqnarray}
where~$\gamma$~is the Lorentz factor of electron. In the exponential gain region, the optical wave extracts energy from electrons and grows exponentially, and meanwhile the average electron energy decreases. The saturation commences when electrons become trapped in the ponderomotive wave and the number of trapped electrons losing energy to the wave is balanced by the electrons gaining energy from the wave. At saturation the so-called sychrotron oscillations are responsible for the development of the sidebands of the radiation spectrum [7].
\subsection{Theoretical estimation}

Here, we consider the FEL process in the sight of longitudinal electron beam phase space (~$\phi ' - \phi$~, where~$\phi$~is the electrons' longitudinal phase). In the exponential gain region, the electrons trapped in the phase space bucket move down from the upper to the bottom to give energy to the radiation wave while the height of the bucket increases with the growth of the radiation power. When the saturation occurs, the height of the bucket almost becomes stable as the radiation power tends to balance. For planar undulator, the bucket height of the longitudinal phase space can be written as
\begin{eqnarray}
H = 2\Omega  = \frac{2}{\gamma }\sqrt {2k_s k_u a_s a_u JJ} 
\end{eqnarray}
Where~$k_u$~,~$k_s$~are the wave numbers of the radiation and undulator field while~$a_u$~,~$a_s$~are the ponderomotive potential of the radiation and undulator field, and~$JJ = J_0 (\xi ) - J_1 (\xi )$~with~$\xi  = {{a_u^2 } \mathord{\left/
 {\vphantom {{a_u^2 } {(2 + 2a_u^2 }}} \right.
 \kern-\nulldelimiterspace} {(2 + 2a_u^2 }})$~. Here, ~$\Omega$~also means the frequency of sychrotron oscillation of the electron phase.
 
At saturation, as the radiation power~$P_{sat.} \approx \rho P_e$~, where~$P_e$~is the power of the electron beam, then we have
\begin{eqnarray}
a_{s,sat.}  = \frac{{2(1 + a_u^2 )}}{{a_u JJ}}\rho ^2 
\end{eqnarray}
Inserting Eq.(3) into Eq.(2), the bucket height at saturation can be estimated as
\begin{eqnarray}
H_{sat.}  = 4\sqrt 2 k_u \rho
\end{eqnarray}

Starting from the position a little before saturation, we use a new undulator with strength parameter of~$K - \Delta K$~to shift the bucket down by the amplitude of~$\Delta \phi ' = H_{sat.}$~to make the radiation wave keep extracting energy from the electron beam, which is equivalent to locating the electrons in the upper of a new bucket. Obviously the resonant electron energy decreases and its variation can be given as
\begin{eqnarray}
\frac{{\Delta \gamma _r }}{{\gamma _r }} = \frac{{\Delta \phi '}}{{2k_u }} = 2\sqrt 2 \rho
\end{eqnarray}
Applying Eq.(1), we obtain the variation of the undulator strength
\begin{eqnarray}
\frac{{\Delta K}}{K} = (1 + \frac{2}{{K^2 }})\frac{{\Delta \gamma _r }}{{\gamma _r }} = 2\sqrt 2 \rho (1 + \frac{2}{{K^2 }})
\end{eqnarray}

However, as the sychrotron oscillation at this time is very fast, the radiation will reach a new saturation soon.

\subsection{Simulation examples}

We use the FEL code GENESIS [8] to simulate SASE FELs with a single-step tapered undulator based on the linac coherent light
source (LCLS) [9] and the Tesla test facility (TTF) [10] like parameters, as shown in Table.1. The start point of the step-tapered
undulator is the last undulator gap before saturation. We have scanned the step-tapered undulator strength with simulations in
steady-state mode. The results are shown in Fig.1 and Fig.2, corresponding to the LCLS and TTF like parameters respectively, which
clearly show that single-step tapering increases the radiation power and the radiation re-saturates in short undulator length.

For LCLS like parameters, the saturation length for normal SASE is 34.08 m that happens in the undulator gap and so the step-tapering
starts from the next undulator segment. From Fig.1, the radiation has the highest saturation power and the shortest saturation length
while ~$\Delta K = 0.375\%$~. According to the theory above, the optimal ~$\Delta K$~ is calculated to be 0.4\%. Similarly, for TTF like
parameters, the optimized ~$\Delta K$~ from simulation is 1.05\% while 1.17\% for theoretical estimation. Obviously the theoretical estimation agrees with the simulation results very well. Furthermore, it is worth to be mentioned that the radiation power has a good tolerance on the step-tapering amplitude.

\begin{center}
\tabcaption{ \label{tab1}  Table.1: GENESIS simulation parameters.}
\footnotesize
\begin{tabular*}{80mm}{c@{\extracolsep{\fill}}ccc}
\toprule Parameter  & LCLS & TTF  \\
\hline
Electron energy /GeV & 4.3 & 1.0 \\
Slice energy spread & $0.025\%$ & $0.02\%$ \\
Peak current /kA & 2.0 & 2.5 \\
Normalized emittance /mm¡¤mrad & 1.2  & 2.0 \\
undulator period /cm & 3.0 & 2.73\\
Undulator strength ~$K$~ & 3.4995 & 1.2671\\
Radiation wavelength /nm & 1.5095 & 6.44578\\
Pierce parameter  & ~$1.22 \times 10^{ - 3}$~ & ~$1.85 \times 10^{ - 3}$~\\
\bottomrule
\end{tabular*}
\end{center}

\begin{center}
\includegraphics[width=8.0cm]{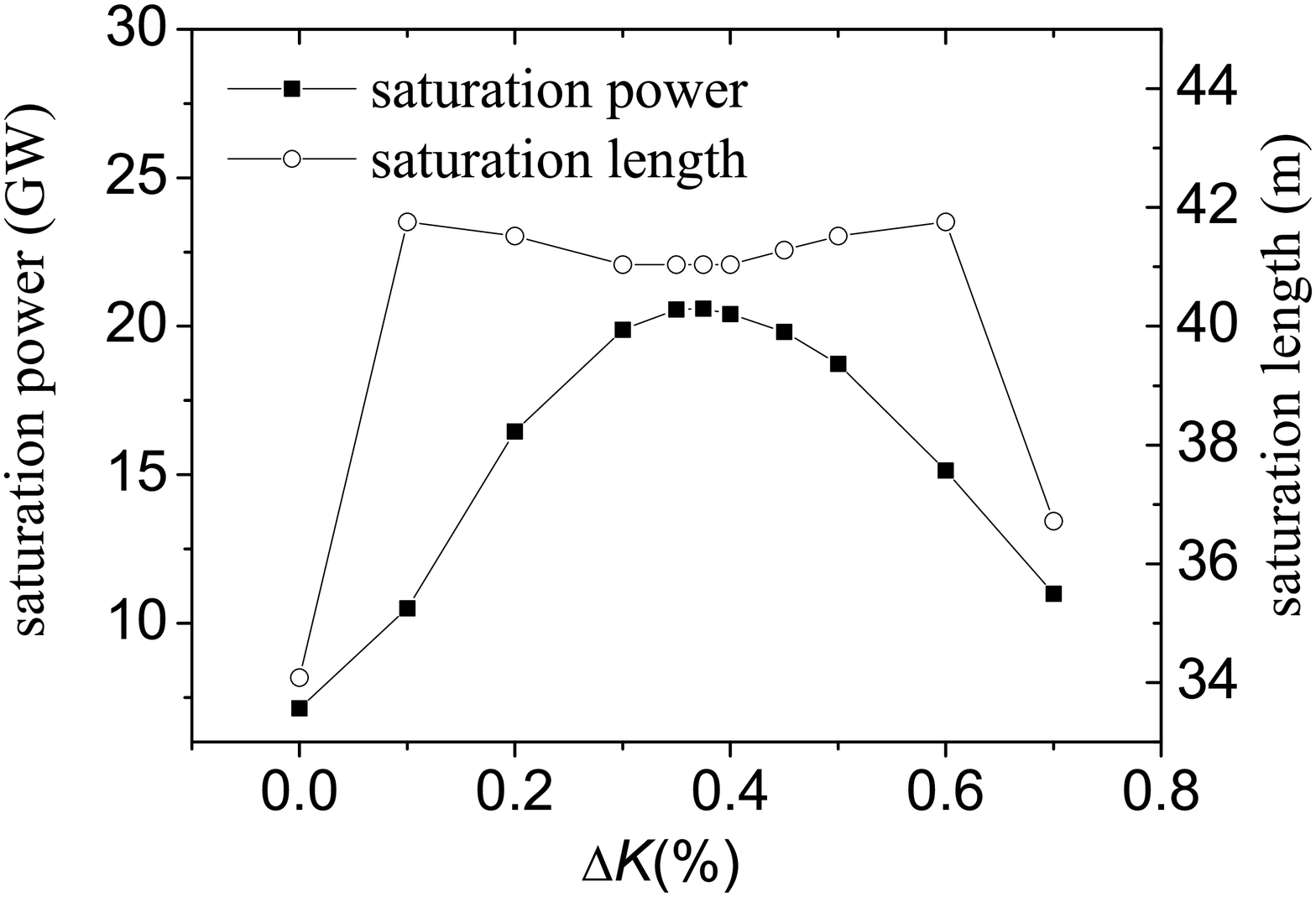}
\figcaption{\label{fig1} The simulation results of saturation power and length with a single-step tapered undulator based on LCLS like parameters. The optimized  for theoretical estimation is 0.4\%.}
\end{center}
\begin{center}
\includegraphics[width=8.0cm]{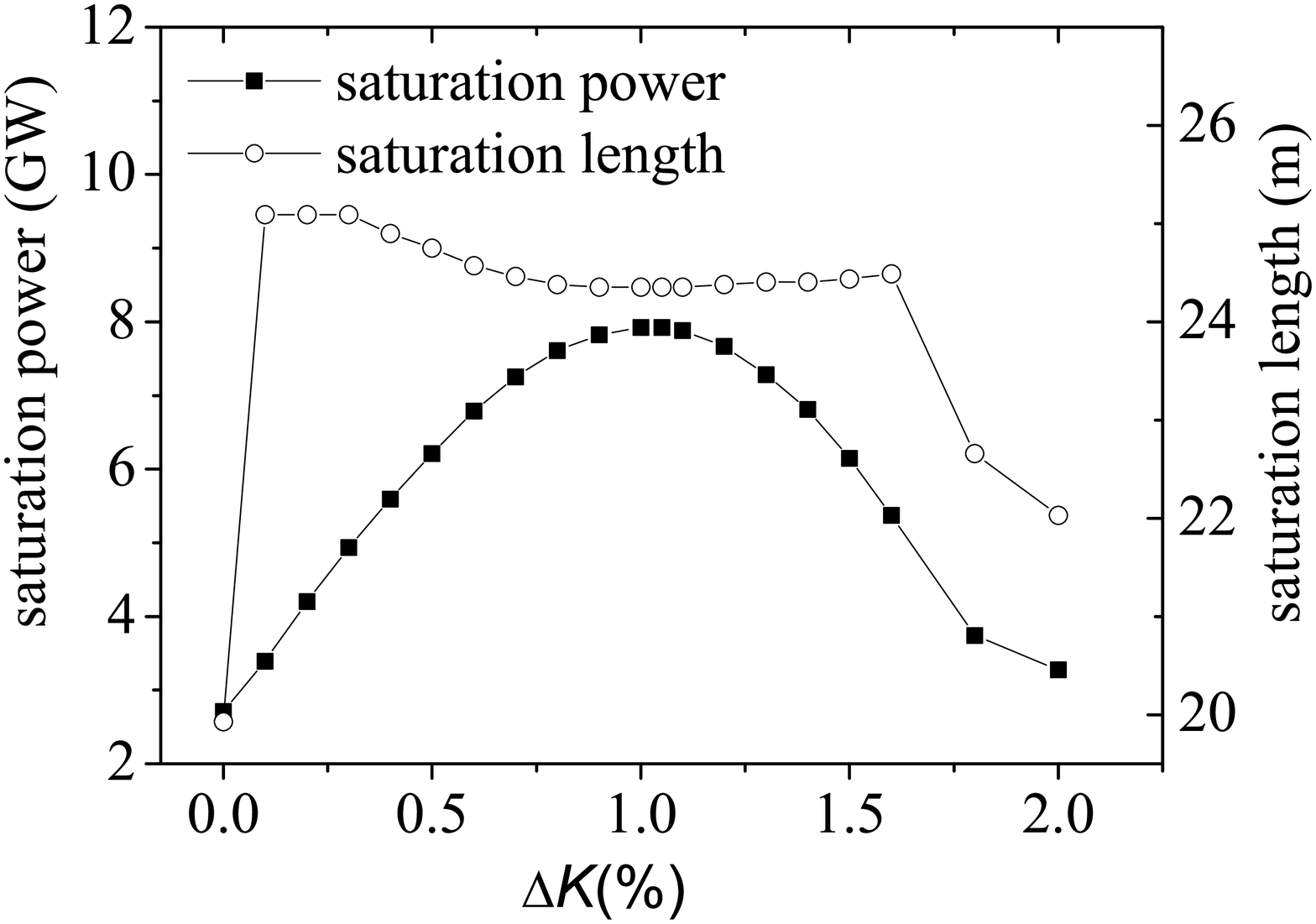}
\figcaption{\label{fig2} The simulation results of saturation power and length with a single-step tapered undulator based on TTF like parameters. The optimized  for theoretical estimation is 1.17\%.}
\end{center}

In addition, we consider starting the single-step tapered undulator from an earlier point. Accordingly, the step-tapering amplitude should be scaled down with the bucket height. The simulation results suggest that the radiation has no stronger power than previous case.

\section{Optimization of energy detuning }
It is well known that using an electron beam with energy above resonance can enhance the radiation power [11]. Same to Part 2, we investigate energy detuning with the longitudinal electron beam phase space. As we know, electrons move down in the bucket with losing energy to the radiation wave, then the bucket height increases and more electrons are captured. According to the conservation of energy,
for a single electron trapped in the bucket, the higher the electron energy is, the more the energy will be exchanged and the stronger
 the saturation power will be. However, for an electron beam with disorganized electron phases, there is a threshold value for the 
 electron energy and when the electron energy exceeds this value and keeps increasing, more and more electrons will not be captured by the bucket. Therefore, the optimal energy detuning should be a balance between the energy loss of single electron and the number of trapped electrons.
 
In high-gain FELs, since the radiation grows exponentially and the bucket height varies very fast, it is difficult to calculate the optimized energy detuning analytically. So we expect to develop empirically optimized energy detuning that maximizes the saturation power through numerical simulations.

We have done many simulations and the results imply that the optimized energy detuning has an approximate linear relationship with the bucket height at normal SASE saturation (~$H_{sat.}$~), as shown in Fig.3. For LCLS like parameters, the electron energy variation corresponding to~$H_{sat.}$~ is~$\Delta \gamma  = {{H_{sat.} } \mathord{\left/
 {\vphantom {{H_{sat.} } {(2k_u )}}} \right.
 \kern-\nulldelimiterspace} {(2k_u )}} = 0.345\% $~while the optimized energy detuning from simulation is~$\delta \gamma _{opt.}  \approx 0.156\%  = 0.452\Delta \gamma$~, where ~$\delta \gamma _{opt.}  = {{(\gamma _{opt.}  - \gamma _r )} \mathord{\left/
 {\vphantom {{(\gamma _{opt.}  - \gamma _r )} {\gamma _r }}} \right.
 \kern-\nulldelimiterspace} {\gamma _r }}$~. And for the case of TTF like parameters,~$\Delta \gamma$~is about 0.523\% while ~$\delta \gamma _{opt.}  \approx 0.235\%  = 0.449\Delta \gamma$~. Based on these, we have the empirical relationship
\begin{eqnarray}
\delta \gamma _{opt.}  \approx {{0.45H_{sat.} } \mathord{\left/
 {\vphantom {{0.45H_{sat.} } {(2k_u )}}} \right.
 \kern-\nulldelimiterspace} {(2k_u )}} = 0.9\sqrt 2 \rho 
\end{eqnarray}

We have checked this empirical formula with other two settings of FEL parameters of FLASH2 and high gain harmonic generation based on Hefei soft x-ray proposal [12], and the results agree with Eq.(7) well.
\begin{center}
\includegraphics[width=8.0cm]{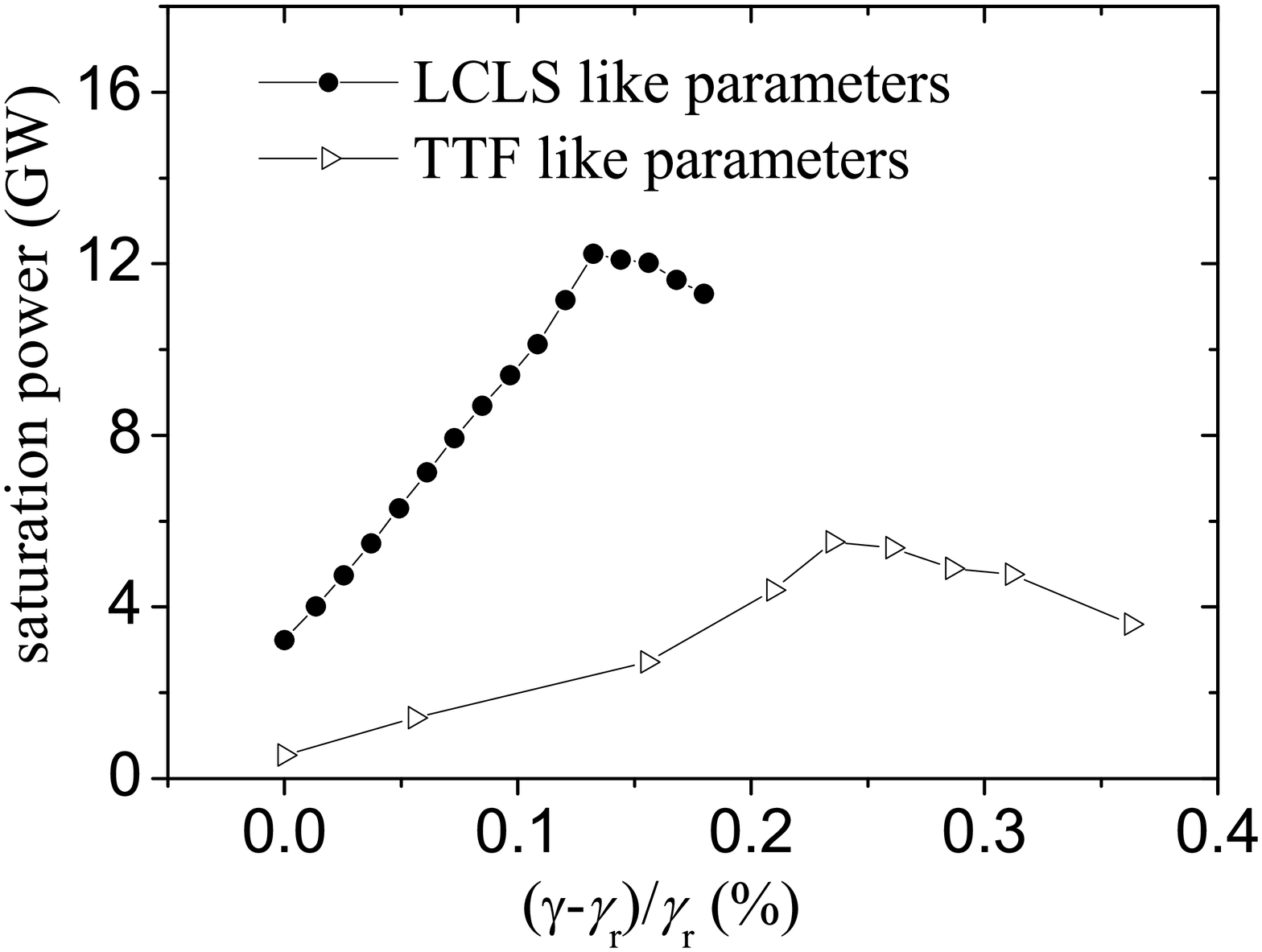}
\figcaption{\label{fig2} The simulation results of saturation power varying with energy detuning based on LCLS and TTF like parameters.}
\end{center}

\section{Conclusions}
In this paper we present a method to estimate the single-step tapering amplitude in the sight of longitudinal electron beam phase space. Through the simulations based on the LCLS and TTF like parameters, we have shown that this method can be an effective way to optimize the undulator parameters in high gain FELs. However, it is especially useful for the FEL facilities whose undulator is just slightly longer than the saturation length.

Furthermore, we have studied the energy detuning and found that the optimized energy detuning is proportionate to the bucket height at normal SASE saturation. An empirical formula has been developed through numerical simulations using the LCLS and TTF like parameters. Then it has been checked with other two FEL parameters settings and the results also agree with the formula well.

These conclusions are effective for high gain FELs, including SASE and seeded FELs.

\end{multicols}

\vspace{-1mm}
\centerline{\rule{80mm}{0.1pt}}
\vspace{2mm}

\begin{multicols}{2}

\end{multicols}
\clearpage

%\end{CJK*}
\end{document}